\documentclass[journal,12pt,draftclsnofoot,onecolumn]{IEEEtran}

\usepackage{amsfonts}
\usepackage{amsmath}

\usepackage{cite}      

\usepackage{graphicx}  

\usepackage{stfloats}  

\usepackage{tablefootnote}

\newcommand{\comment}[1]{}
\newcommand{\mycomment}[1]{}
\newcommand{\gmcomment}[1]{}
\newcommand{\obmcmntout}[1]{} %
\newcommand{\obmreuse}[1]{} %
\newcommand{\obmsubmittodo}[1]{} %
\newcommand{\psbfg}[1]{} %
\newcommand{\omgc}[1]{} %

\newcommand{\obmabrdgol}[1]{} %
\newcommand{\obmthg}[1]{} 

\newcommand{\figcomment}[1]{}

\begin{document}

\title{Quantified Spectrum Sharing:\\ Motivation, Approach, and Benefits}


\author{Nilesh~Khambekar,~\IEEEmembership{Member,~IEEE,}
        ~Chad~M.~Spooner,~\IEEEmembership{Senior~Member,~IEEE,}
        and~Vipin~Chaudhary,~\IEEEmembership{Member,~IEEE} 
\thanks{Nilesh Khambekar and Vipin Chaudhary are with the Department of Computer Science and Engineering, University at Buffalo, SUNY, Buffalo, New York 14260-2000. \{nvk3, vipin\}@buffalo.edu}
\thanks{Chad M. Spooner is with NorthWest Research Associates, Monterey, CA. cmspooner@nwra.com}}


%


\maketitle


\begin{abstract}
A significant portion of the radio frequency spectrum remains underutilized with exclusive and static allocation of spectrum. The growing demand for spectrum has spurred a need for dynamic spectrum sharing paradigm. While the new dynamic spectrum sharing paradigm helps to improve utilization of the precious spectrum resource, there exist several  obstacles on the technical, regulatory, and business fronts for the adoption of the new paradigm. 

In this paper, we investigate the limitations of the existing techniques and argue for quantified approach to dynamic spectrum sharing and management. We introduce a quantified approach to spectrum sharing based on defining and enforcing quantified spectrum-access rights. By discretizing the spectrum-space in the time, space, frequency dimensions, this approach enables quantifying the spectrum consumed by individual transceivers. It enables defining and enforcing a quantified spectrum-access policy in real-time. The proposed quantified approach brings in simplicity, precision and efficiency in terms of spectrum commerce and operations while addressing the key technical and regulatory challenges.

\end{abstract}



%
\IEEEpeerreviewmaketitle


\setlength{\textfloatsep}{5pt}

\section{Introduction}

Traditionally, the radio frequency (RF) spectrum has been statically and exclusively allocated to wireless services. This static spectrum allocation paradigm results into inefficient usage of the spectrum in the time, space, and frequency dimensions\cite{fcc_sptf, ssc_erpek}. In order to meet the growing demand for the new and high bandwidth wireless services, the RF spectrum needs to be dynamically shared by multiple wireless service providers \cite{fcc_nprm_2003, pcast, dod_ems}. 

The dynamic spectrum sharing paradigm brings in new challenges on technical, regulatory, and business fronts; for example, it is not trivial to understand \textit{how much} spectrum is available for sharing in the time, space, and frequency dimensions. For effective spectrum sharing, non-harmful interference needs to be ensured to the receivers in the system given the non-deterministic propagation and dynamic spectrum-access conditions. Due to the aggregate interference effects, dynamic propagation conditions, and software defined capabilities, the regulation of dynamic spectrum-access constraints is a complex issue. Furthermore, from a business perspective, it is also important to be able to flexibly and efficiently share or trade the spectrum in addition to solving the core technical and regulatory difficulties.

In this paper, we investigate the underlying issues and uncover the limitations of the existing spectrum sharing approaches. We argue for the need of a quantified approach to spectrum sharing. We propose to divide the spectrum-space into multiple \textit{unit-spectrum-spaces} and quantify the use of spectrum by each of the individual transceivers in the space, time, and frequency dimensions. The proposed \textit{discretization and quantification of spectrum-space} (DQSS) approach facilitates dynamically defining, controlling, and enforcing spectrum-access rights.




The rest of this paper is organized as follows. In Section II, we illustrate the limitations of the existing spectrum sharing technologies and underscore the need for a quantified approach to spectrum sharing. In Section III, we present the quantified approach, DQSS, to dynamic spectrum sharing based on spectrum-space discretization and quantification of the use of spectrum by individual transceivers. We further extend the quantitative approach to spectrum management, spectrum operations, spectrum regulations, and spectrum commerce. In Section IV, we discuss the benefits of quantified approach in terms of addressing the challenges for adopting the dynamic spectrum sharing paradigm. Finally, in Section V, we draw conclusions and outline further research avenues. 

\section{Motivation}

The dynamic spectrum sharing approaches have been evolving since the past decade \cite{dsa_survey, akyildiz_survey1, winn_survey, fcc_cbrs}. Depending on the degree of sharing, the various spectrum sharing approaches fall into exclusive spectrum use, static spectrum sharing, dynamic spectrum sharing, and pure spectrum sharing categories \cite{winn_survey}. However, in terms of articulating the spectrum-access rights, the spectrum sharing models primarily resort to statically or dynamically defining a spatio-temporal boundary along with a \textit{fixed} set of constraints. This not only undermines the potential for maximizing the use of spectrum but also leads to several technical, regulatory, operational, and business difficulties. An example is the limited success story of dynamic spectrum sharing in UHF bands.

In Nov. 2008, Federal Communications Commission (FCC) released a Notice of Proposed Rule Making (NPRM) to allow the unlicensed radios to operate in the TV bands without causing harmful interference to the incumbent services \cite{fcc_nprm_2008}. The Opportunistic Spectrum Access (OSA) of the unused UHF bands received a wide commercial interest for several potential wireless services; However, the performance estimation studies of OSA have revealed that the amount of the \textit{implied} available spectrum is \textit{very limited} to meet the increasing demand for RF spectrum \cite{smm_thesis, berk_wsc, osa_feasib}. Moreover, the secondary users cannot ensure desired quality of service necessary for the business cases due to the \textit{secondary} rights for accessing the spectrum. On the other hand, incumbents do not have any incentive for sharing the spectrum. Furthermore, the secondary access to the spectrum is very hard to regulate. Considering interference aggregation effects, dynamic nature of propagation conditions, and dynamic spectrum-access scenarios, the primary owners of the spectrum need a way to confirm that their receivers are not subjected to harmful interference and the service experience is not degraded. This requires the ability to reliably estimate the interference margin at the receivers and accordingly infer the maximum transmit-power at the secondary transmitter positions. Furthermore, the behavior of software defined radio devices could be altered with software changes and thus the service is exposed to attacks from the secondary users of the spectrum. In order to ensure protection of the spectrum rights, the spectrum-access constraints need to be \textit{enforceable}. 

We observe that the decisions for exercising spectrum-access in case of OSA are based on detection of primary transmitter signal using a certain specified radio sensitivity. In this case, the decision for spectrum-access is binary in nature. This gives rise to `\textit{not enough spectrum for secondary usage}' if the policy for shared spectrum-access is conservative and `\textit{no guarantee for ensuring service quality}' if the shared spectrum-access policy is aggressive. The binary nature of the spectrum-access decision cannot protect the spectrum rights of incumbents and requires the spectrum-access policy to be increasingly conservative to guard against interference aggregation. Therefore, when multiple secondary transmitters exercise spectrum-access, we need to \textit{quantitatively articulate the spectrum-access rights}. This helps maximizing a spectrum-access opportunity without causing harmful interference. If technical and regulatory problems are solved, more and more incumbents will have an incentive to share the spatially, temporally, and spectrally unexploited spectrum. 

\begin{figure*}[htbp!]
\centering
{\includegraphics [width=0.92\textwidth, angle=0] {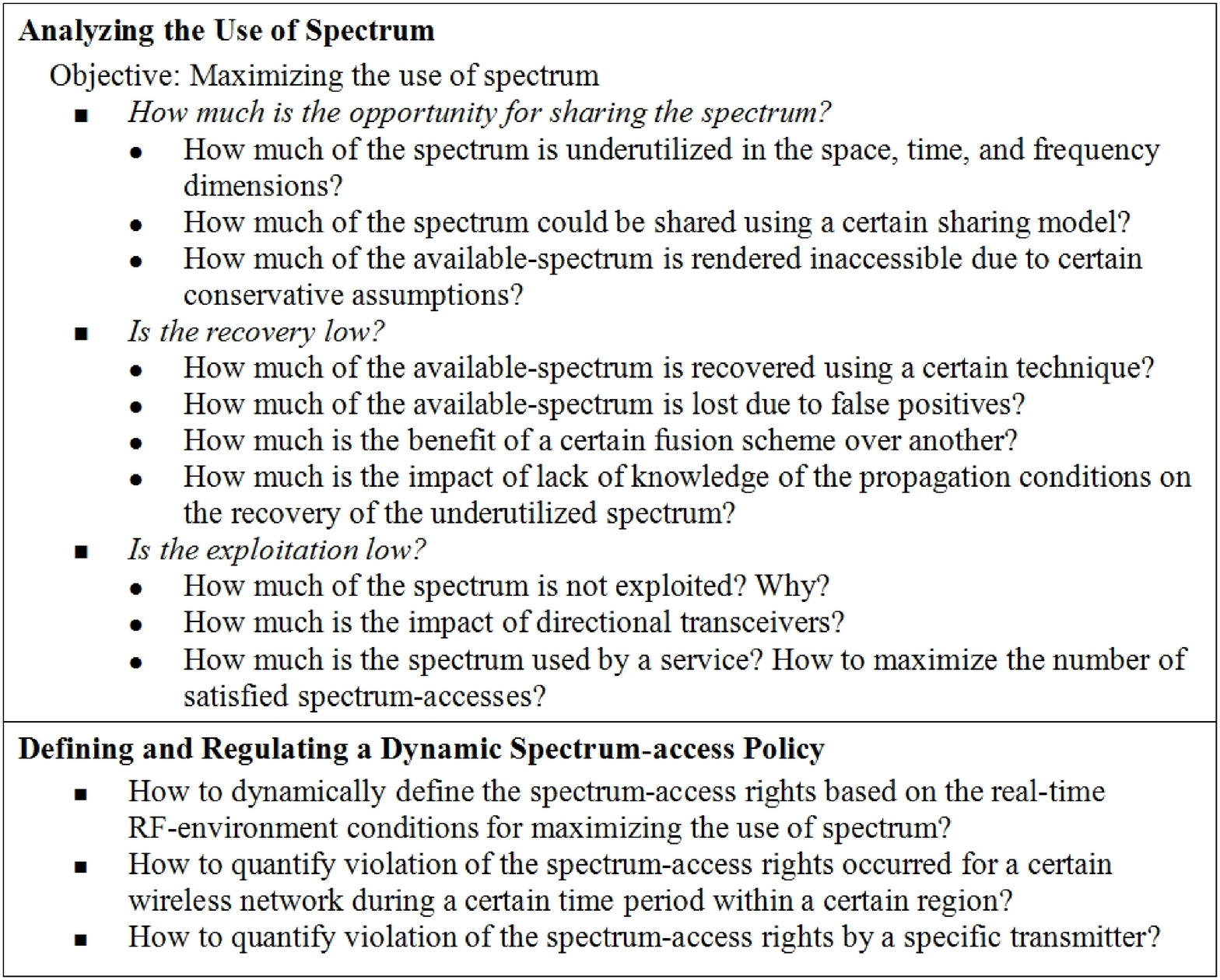}}
\setlength{\abovecaptionskip}{-3pt}
\caption{Example questions in case of optimizing a typical dynamic spectrum sharing scenario. The questions shade light on the various quantitative decisions involved with regards to spectrum sharing and spectrum management. The question-map emphasizes on the need for a methodology to characterize and quantify the use of spectrum in order to effectively manage the use of spectrum.}
\label{fig:qdsa_qneed}
\end{figure*}

Figure~\ref{fig:qdsa_qneed} illustrates the need for a methodology to characterize and quantify the use of spectrum under dynamic spectrum sharing paradigm with the aid of a question-map. The question-map enumerates the quantitative decisions involved in the process of investigating the weaknesses of a spectrum sharing mechanism, comparing various algorithms and architectures for recovery and exploitation of the spectrum, and optimizing the spectrum sharing opportunities. 

Traditionally the performance of spectrum recovery is measured in terms of the throughput for the secondary users and outage probability \cite{mishra_coop_sensing, ganesan, Visotsky}. The performance of detection of spectrum holes is also captured in terms of probability of missed detection and false positives \cite{ganesan_spthole, raman, tuando}. However, this characterization of the performance is in the context of spectrum sharing constraints defined by a certain spectrum sharing model or in terms of system-level objectives. In order to maximize the use of spectrum, we need a methodology that can characterize the performance of the recovery and exploitation of the underutilized spectrum in the space, time, and frequency dimensions. 

The existing methodologies to define the use of spectrum and quantify its efficiency are based on the static spectrum assignment paradigm and are not suitable for the dynamic spectrum sharing paradigm. ITU defined \textit{spectrum utilization factor} as product of the frequency bandwidth, geometric space, and the time denied to other potential users \cite{itumetrics}. However, spectrum utilization factor does not represent \textit{actual usage}. For example, if a licensed user does not perform any transmissions, the spectrum is still considered to be \textit{used}. It also cannot quantify the use of spectrum under spatial overlap of wireless services. The IEEE 1900.5.2 draft standard captures spectrum usage in terms of transceiver-model parameters and applies standard methods for ensuring compatibility between the spectrum sharing networks \cite{ieee_compat}. Thus, the approach helps to ensure compatibility; however, it cannot \textit{characterize and quantify} the use of spectrum and the performance of spectrum management functions.

Finally, from a business perspective, the ability to qualitatively and quantitatively interpret a spectrum sharing opportunity in a certain frequency band within a geographical region of interest is essential in order to evaluate its business potential. With the change in paradigm, businesses need the ability to control the use of spectrum at a fine granularity in order to maximize fine granular spectrum-reuse opportunities. With spectrum as a quantified resource perspective, the spectrum trade conversation could be on the following lines: \textit{``I have `x' units of spectrum right now, I have given `y' units of spectrum to somebody and have `z' units of spare spectrum which I can share".} Also, the quantification of the use of spectrum would provide insight into the business implications of a dynamically identified spectrum-access opportunity in terms of the service quality, range, and user experience. 


\section{DQSS: Quantified Approach to Dynamic Spectrum Management, Operations, and Regulations}

The proposed spectrum sharing approach is based on the MUSE \textit{M}ethodology to articulate, characterize, and quantify the \textit{USE} of spectrum in the space, time, and frequency dimensions. MUSE is presented in \cite{ong_qsh2}. Here, we first describe what constitutes use of the spectrum. This helps us to define the system model for quantified dynamic spectrum sharing.

\subsection{How is spectrum consumed?}

Traditionally, we assume that spectrum is consumed by transmitters; however, the spectrum is \textit{also} consumed by receivers by constraining spectrum access by other transmitters. We note that for guaranteeing successful reception, protection is traditionally accomplished in term of guard-bands, separation distance, and constraints on operational hours. Thus, the presence of receivers enforces limits on the interference-power in the space, time, and frequency dimensions. When the access to spectrum is exclusive in the time, space, and frequency dimensions, the spectrum consumption by receivers need not be separately considered \cite{itu}.

Understanding that spectrum is consumed at the granularity of an individual transceiver, helps us define the granularity of spectrum-access. Under quantified dynamic spectrum sharing approach, we consider multiple heterogeneous spatially-overlapping wireless networks sharing spectrum in the time, space, and frequency dimensions. In order to facilitate the ability of defining and enforcing spectrum-access footprints at the granularity of a transceiver, we seek to capture spectrum-access at the \textit{lowest granularity of spectrum consumption} in the system, that is, at the granularity of an individual transceiver. 

\subsection{System Model}
\begin{figure}[htbp!]
\centering
{\includegraphics [width=0.48\textwidth, angle=0] {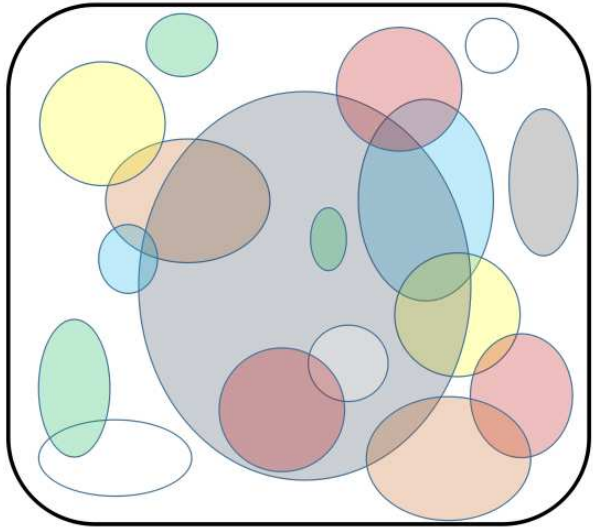}}
\setlength{\abovecaptionskip}{-4pt}
\caption{System model with multiple spatially overlapping heterogeneous RF-systems sharing spectrum in the space, time, and frequency dimensions.}
\label{fig:gensys}
\end{figure}

We consider a generic system with multiple heterogeneous wireless services sharing the RF-spectrum.  We define a \textit{RF-link} represents a transmitter and one or more receivers exercising spectrum-access. A \textit{RF-network} represents an aggregate of RF-links. We refer to the aggregate of RF-networks sharing a spectrum space in the time, space, and frequency dimensions within the geographical region of interest as a \text{RF-system} and such multiple RF-systems are sharing the spectrum in the time, space, and frequency dimensions within the geographical region of interest. 

Under the system model, we consider that the transceivers optionally employ directional transmission and reception in order to minimize interference. A receiver can withstand certain interference when the received Signal to Interference and Noise Ratio (SINR) is greater than a receiver-specific threshold, $\beta$\footnote{The threshold, $\beta$, represents the quality of a receiver and incorporates receiver-noise and other receiver technology imperfections. Thus, $\beta$ models the receiver-performance under quantified dynamic spectrum sharing model.}. 

Let $P_{MAX}$ represent the maximum permissible power at any point and $P_{MIN}$ represent arbitrary minimum power at any point in the system. $P_{MIN}$ could be chosen to be a very low value below the thermal noise floor. The difference between the maximum and minimum spectrum consumption at a point represents the spectrum consumed by a transceiver at a point. Thus, the maximum spectrum consumption at a point is given by
\begin{equation}
P_{CMAX} = P_{MAX} - P_{MIN}	
\end{equation}
  
\subsection{Overview of the MUSE Methodology}


\noindent
\textbf{Spectrum-space discretization} 
The spectrum consumed by a transmitter or a receiver is continuous in the space, time, and frequency dimensions. In order to quantify spectrum consumption, we discretize the total spectrum-space within a geographical region of interest in the space, time, and frequency dimensions.

\noindent
\textbf{A unit spectrum-space}
\noindent
A unit spectrum-space represents the spectrum within an unit area, in a unit time, and a unit frequency band. A unit spectrum-space thus represents the lowest granularity of spectrum sharing. 

\noindent
\textbf{The total spectrum-space}
\noindent
Let the geographical region be discretized into $\hat{A}$ unit-regions, $\hat{B}$ unit-frequency-bands, and $\hat{T}$ unit-time-quanta.  Thus, the total spectrum-space is given by
\begin{equation}
\label{eq:p1totalspectrum}
\Psi_{Total} = P_{CMAX} \hat{T} \hat{A} \hat{B} .
\end{equation}
The unit of the total spectrum is $Wm^2$. 

\noindent
\textbf{A spectrum consumption space:}
\noindent
A spectrum consumption space captures the spectrum consumption by an entity in the discretized spectrum-space. The entity could be an individual transceiver or a collection of transceivers. 

\noindent
\textbf{Quantification of a spectrum consumption space:} 
\noindent
In order to quantify the spectrum consumption by an entity over a range of time, space, and frequency, borrowing the classic discretization principle \cite{dsp_prkpr}, we \textit{sample} the spectrum-consumption by the individual transceivers in the time, space, and frequency dimensions and aggregate the discretized spectrum consumption across all the unit spectrum-spaces within the geographical region of interest. 

\noindent
\textbf{Quantification of available spectrum space:} 
\noindent
Quantifying the spectrum consumed by all the transmitters and receivers in a unit spectrum-space enables quantifying \textit{unit spectrum space opportunity}, that is the unconsumed spectrum in the unit spectrum-space. The amount of available spectrum spectrum within a geographical region is obtained by aggregating the unit spectrum-space opportunity across all the unit spectrum-spaces.




\subsection{Illustration: Use of the Spectrum in a Geographical Region}
\begin{figure}[ht]
\begin{center}
{\includegraphics [width=0.64\textwidth, angle=0] {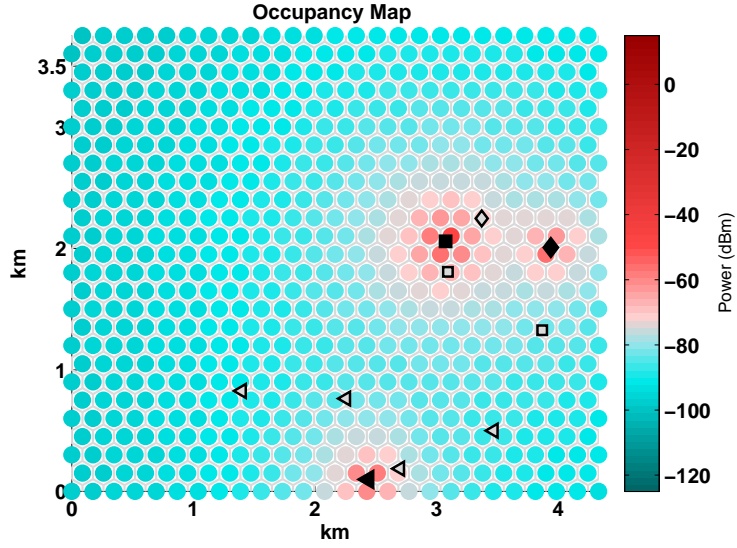}}
\setlength{\abovecaptionskip}{-12pt}
\setlength{\belowcaptionskip}{-24pt}
\caption{Single-band spectrum-occupancy map showing the aggregate RF power across the unit-regions within a geographical region. Transmitters and receivers in a single network have the same shape; transmitter is solid.}
\label{figure:occupancy_map}
\end{center}
\end{figure}

\begin{figure}[ht]
\begin{center}
{\includegraphics [width=0.64\textwidth, angle=0] {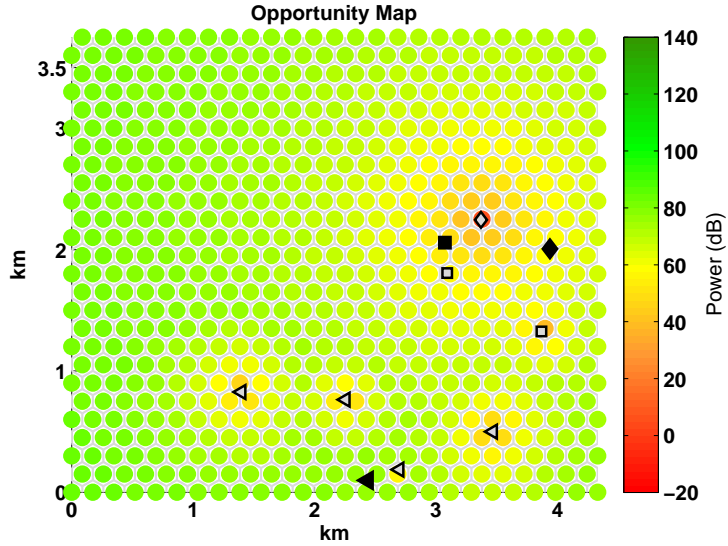}}
\setlength{\abovecaptionskip}{-12pt}
\setlength{\belowcaptionskip}{-24pt}
\caption{Single-band spectrum-opportunity map showing the RF power that each unit-region can tolerate given the presence of the shown networks. High-opportunity regions are green; low are red. The spectrum opportunity is relative to -125 dBm ($P_{MIN}$).}
\setlength{\belowcaptionskip}{-16pt}
\label{figure:opportunity_map}
\end{center}
\end{figure}

By characterizing the \textit{spectrum-occupancy} across the unit-spectrum-spaces within a geographical region, we can identify the \textit{utilized spectrum}. Similarly, by characterizing the \textit{spectrum-opportunity} across the unit-spectrum-spaces, we can identify the \textit{available spectrum}. Figures~\ref{figure:occupancy_map} and \ref{figure:opportunity_map} capture the spatial distribution of spectrum occupancy and spectrum opportunity respectively.

\subsection{Applying Quantified Approach}

\begin{figure}[htbp!]
\centering
{\includegraphics [width=0.64\textwidth, angle=0] {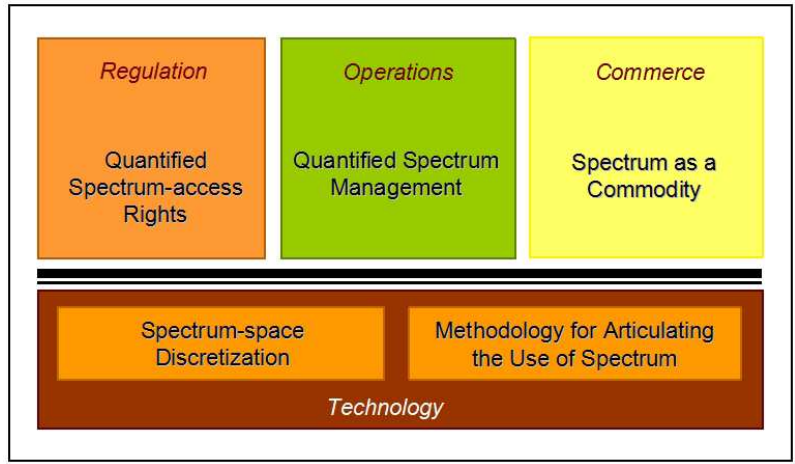}}
\setlength{\abovecaptionskip}{-8pt}
\caption{MUSE and Spectrum-space discretization enable quantified approach and help to address the challenges in spectrum regulation, spectrum operations, and spectrum commerce.}
\label{fig:qss}
\end{figure}

Spectrum-space discretization and MUSE define the foundation of our quantified approach. Here, we extend the quantified approach to key use-cases such as dynamic spectrum-access, spectrum operations, and spectrum-commerce. Figure~\ref{fig:qss} depicts quantified DSA, quantified spectrum sharing, and quantified spectrum management techniques enabled by the underlying Spectrum-space discretization technique and MUSE methodology.

\subsubsection{Quantified DSA}
Under Quantified DSA, the spectrum-rights are defined in terms of  the amount of spectrum consumed by the transceivers. Thus, Quantified DSA is independent of the spectrum-sharing model applied.  Quantification of use of spectrum by individual transceiver helps to address several challenges in spectrum regulation. For example, it is possible to quantify the harmful interference caused by a certain transmitter when the receiver is subjected to aggregate harmful interference. Quantified spectrum-rights imply that operators do not need to impose spatio-temporal  boundaries and can accomplish cohabitation of multiple spatially-overlapping heterogeneous RF-systems. 

\subsubsection{Quantified Spectrum Management}
With dynamic spectrum sharing paradigm, we identify two core spectrum management functions: \textit{producing} spectrum-resource and \textit{consuming} spectrum-resource. In a broader context, considering the various spectrum sharing approaches, we loosely call these two functions as spectrum harvesting and spectrum exploitation respectively.

\noindent
\textbf{Quantified Spectrum Harvesting}
Traditionally underutilized spectrum is identified by detection of transmitter signal. In this case, the performance of spectrum sensing is measured in terms of probability of missed detection and false positives \cite{liang}. Quantified spectrum harvesting senses the RF-environment in order to estimate unit-spectrum-space opportunity and thus enables to choose optimum spectrum-access parameters. One of the key benefits of quantified approach to spectrum harvesting is aggregation of unit-spectrum-space opportunities across multiple frequency bands. \textit{Spectrum aggregation} helps better scheduling of spectrum-access requests and enables advanced routing of the spectrum connections. 

\noindent
\textbf{Quantified Spectrum Exploitation}
When quantified spectrum-access footprints are assigned to RF-entities, it enables us to understand how efficiently the harvested spectrum is exploited.  It also enables us to quantify the amount of harmful interference caused to the cochannel receivers by an individual transmitter or a collection of transmitters. Thus, appropriate spectrum assignment schemes could be developed that can \textit{quantitatively} control the spectrum-footprint assigned to each of the transmitters. With a quantified approach to spectrum consumption, the traditional spectrum-scheduling and spectrum-allocation problems are transformed into a problem of optimizing the spectrum consumption spaces for a set of spectrum-access requests \cite{oms3_sco}. This enables us to efficiently exploit the harvested available spectrum. 

\subsubsection{Quantified Spectrum Commerce}
Discretization of the spectrum-space and quantification of the usage in the discretized spectrum-space enables us to treat spectrum as a commodity. The estimated quantity of available-spectrum determines the supply of the commodity. With quantified spectrum commerce, the pricing is defined based on the quantify of the spectrum consumed. With the ability to quantitatively understand the spectrum consumed by a single transceiver, a single RF-system, an aggregate of RF-systems, this pricing can be very precise. 

\subsection{Defining and Enforcing a Quantified Spectrum Access Policy}
\begin{figure}[htbp!]
\centering
{\includegraphics [width=0.64\textwidth, angle=0] {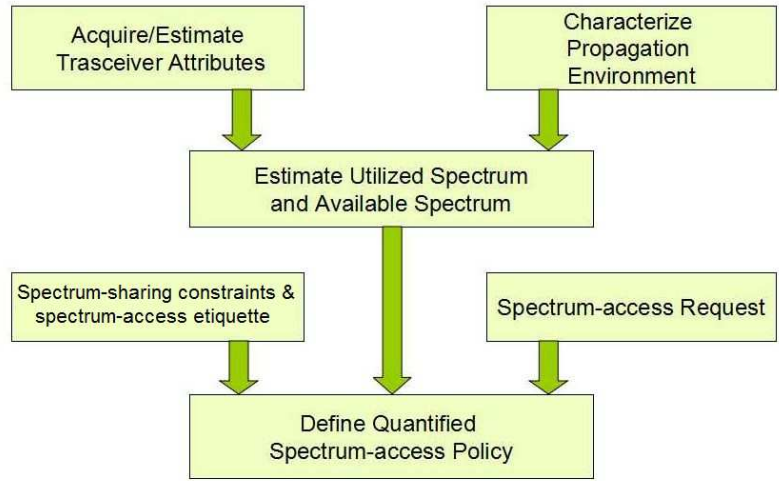}}
\caption{Illustration of defining spectrum-access rights based on real-time use of the spectrum. By passively estimating the spectrum-access attributes of the transmitters and by characterizing the propagation environment, the use of spectrum by transmitters and the available spectrum can be estimated. Based on spectrum-sharing constraints and etiquette, quantified spectrum-access rights can be defined and enforced in real time.}
\label{fig:ill_qdsap}
\end{figure}

Estimation of the use of spectrum by each of the transceivers in the space, time, and frequency dimension enables defining and enforcing quantified spectrum-access rights. Figure~\ref{fig:ill_qdsap} describes the overall approach for defining a policy with quantified spectrum-access rights in real time. A spectrum-sharing model may choose to add a guard-margin to the estimated spectrum-opportunity. A spectrum-access mechanism (SAM) may further control the spectrum consumed by the to-be-added transceivers and thereby increase the overall number of spectrum accesses. Thus, the spectrum-access rights for the transceivers are defined based on the real-time spectrum-access opportunity, spectrum-sharing constraints, and spectrum-access etiquette. The rights are articulated in terms of allowed use of the spectrum in the space, time, and frequency dimensions and accordingly spectrum-access parameters for the transceivers can be inferred. 

In \cite{oms4_cf1}, by using an external dedicated RF-sensor network and cochannel interference tolerant signal processing techniques, we illustrate estimating the available spectrum and transceiver-utilized spectrum.

\section{Benefits Towards Addressing the Challenges for the Dynamic Spectrum Sharing Paradigm} 
In this section, we enumerate benefits of DQSS from technical, operational, regulatory, and business perspectives. 
\subsection{Benefits Towards Spectrum Management}
\begin{itemize}
  \item DQSS helps to characterize and quantify the use of spectrum at the desired granularity in the space, time, and frequency dimensions. MUSE helps to query how much spectrum is consumed by a single transceiver or any logical collection of the transceivers.   
	\item DQSS helps to compare, analyze, and optimize the performance of spectrum management functions. For example, it is possible to quantitatively analyze performance of ability to recover the underutilized spectrum of various spectrum sensing algorithms (like energy-detection, cyclostationary feature detection) or various cooperative spectrum sensing infrastructures  based on the recovered spectrum space, lost-available spectrum space, and potentially-incursed spectrum space. 
	\item DQSS can help to estimate the available spectrum and the exploited spectrum. Thus, it offers the ability to define the spectrum-access rights based on the real-time RF-environment conditions. Using the real-time RF-environment conditions helps to get rid of conservative assumptions and make an efficient use of the spectrum.
	\item The proposed spectrum-discretization approach facilitates adaptation of the spectrum management functions under dynamic RF environment conditions and dynamic spectrum-access scenarios.  
\end{itemize}

\subsection{Benefits Towards Dynamic Spectrum Access}
\begin{itemize}
  \item DQSS enables us to articulate, define, and enforce spectrum-access rights in terms of the use of spectrum by the individual transceivers.
	\item From operations perspective, the guard space could be effectively controlled. The discretized spectrum management approach enables us to easily map a guard margin value to the amount of the inexercisable spectrum. Thus, depending on the user-scenario, spectrum sharing behavior could be changed with visibility into the implied availability of the spectrum.
	\item Another advantage from an operational perspective is controlling the granularity of spectrum sharing. With discretized approach to spectrum management, the dimensions of a unit-spectrum-space imply the granularity of sharing of the spectrum resource. 
	\item With characterization of spectrum-access opportunity in the space, time, and frequency, DQSS provides the ability to share spectrum without defining a boundary across spectrum uses. 
	\item The discretized spectrum management can be applied independent of the spectrum sharing model. Thus, it can be applied in case of the completely dynamic spectrum sharing model like pure spectrum sharing model or even in case of a conservative spectrum sharing model like static spectrum sharing model.
	\item From a regulatory perspective, DQSS offers the ability to enforce a spectrum-access policy and ensure protection of the spectrum rights of the users. As the spectrum-access rights are identified at the granularity of a single transceiver, the violations by a particular transmitter, or the harmful interference for the individual receivers could be characterized and quantified.
\end{itemize}

\subsection{Benefits Towards Spectrum Commerce}
\begin{itemize}
	\item The quantified approach brings in simplicity in spectrum trade. It enables easier understanding and interpretation of the outcomes; thus, it requires less skills of its users. 
	\item The quantified approach enables to investigate the amount of the spectrum that can shared and evaluate the potential for a business opportunity.
	\item From a business development perspective, spectrum sharing models devised using a quantified approach enable spatial overlap of multiple RF-systems and avoid spatial fragmentation of coverage. This is important for defining new services exercising shared spectrum-access rights. 
	\item Aggregation of fine granular spectrum sharing opportunities gives incentives for spectrum-owners to extract more value out of their underutilized spectrum; a bigger spectrum-pool is attractive for secondary users as well. Thus, characterization of the fine granular spectrum-access opportunities enables building a bigger spectrum-resource pool. 
\end{itemize}

\section{Conclusions and Future Research Avenues}


With the static and exclusive spectrum allocation paradigm, the spectrum usage by the receivers need not be explicitly considered. Under the new dynamic spectrum sharing paradigm, multiple spatially-overlapping heterogeneous wireless networks exercise a shared access to the spectrum. This necessitates considering the spectrum used by the individual transmitters \textit{and} receivers.

\textit{Spectrum-space discretization} facilitates quantifying the use of spectrum and enables quantified spectrum sharing and management. Quantification of the use of spectrum enables to treat spectrum as a commodity and brings in simplicity, precision, and efficiency into the business models under the new dynamic spectrum sharing paradigm.  

\textit{Quantified} DSA makes it possible to precisely control the use of spectrum under spectrum sharing and ensure non-harmful interference among \textit{all} spectrum sharing networks. Defining and enforcing spectrum-access rights in a \textit{quantified} manner enables us to maximize the utilization and availability of spectrum and accomplish efficient utilization of the limited resource.

We encourage defining and enforcing spectrum access rights in real-time. Although this requires a dedicated spectrum management infrastructure, it potentially brings in new business models along with flexible and efficient use of the spectrum and an ability for automated regulation of the dynamic spectrum-accesses. 

We note that the dynamic spectrum sharing paradigm calls for fundamental changes in spectrum management functions. The proposed quantified approach brings in advanced spectrum management functions such as quantified spectrum harvesting, quantified spectrum exploitation, spectrum aggregation, advanced spectrum routing, fine-grained spectrum management, and cognitive spectrum management. These are some examples of the research problems offered by quantified dynamic spectrum sharing that will help to realize the potential of dynamic spectrum sharing paradigm. 





\bibliographystyle{IEEEtran}


\mycomment{

\bibliographystyle{IEEEtran}

}


\end{document}